\input amstex
\documentstyle{amsppt}

\def\L{\operatorname{L}}
\def\Id{\operatorname{Id}}
\def\B{\operatorname{B}}
\def\diag{\operatorname{diag}}
\def\S{\operatorname{S}}
\def\rank{\operatorname{rank}}

\document
\hyphenation{Lo-ba-chev-sky spa-ce}

\centerline{\bf ON ELLIPTICAL BILLIARDS IN THE LOBACHEVSKY SPACE AND}
\centerline{\bf ASSOCIATED  GEODESIC HIERARCHIES}

\

\centerline
{\smc Vladimir Dragovi\' c\footnote{vladad\@mi.sanu.ac.yu, on leave at SISSA, Via Beirut 2-4, Trieste, Italy},
Bo\v zidar Jovanovi\' c\footnote{bozaj\@mi.sanu.ac.yu},
Milena Radnovi\' c\footnote{milena\@mi.sanu.ac.yu}}

\

\centerline{Mathematical Institute SANU}
\centerline{Kneza Mihaila 35, 11000 Belgrade}
\centerline{Serbia, Yugoslavia}

\

\centerline{\smc Abstract}

\

We derive Cayley's type
conditions for periodical trajectories
for the billiard within an ellipsoid in the
Lobachevsky space. It appears that these
new conditions are of the same form as those obtained before for
the Euclidean case.
We explain this coincidence 
by using theory of geodesically equivalent
metrics and show that Lobachevsky and Euclidean elliptic billiards can be
naturally considered as a part of a hierarchy of integrable
elliptical billiards.


{\it Subj.  Class. Geometry of integrable systems}


\bigskip

\head {1. Introduction}
\endhead

\medskip

We start with the following well-known
integrable mechanical system: motion
of a free particle within an ellipsoid in the Euclidean space of any
dimension $d$. On the boundary, the particle obeys the billiard law.
Integrability of the system is related to classical
geometrical properties of elliptical billiards: the Chasles, Poncelet
and Cayley theorems.
According to the Chasles theorem [1] every line in this space is tangent
to $d-1$ quadrics confocal to the outer ellipsoid. Even more, all segments
of the particle's trajectory are tangent to the same $d-1$ quadrics [26].
The Poncelet theorem [28, 22, 13] put some light on closed
billiard trajectories: {\it there exists a closed trajectory with $d-1$ given
confocal caustics if and only if infinitely many such trajectories exist,
 and all of them have the same period.} Since the periodicity of a billiard
 trajectory depends only on its caustic surfaces, it is a natural question
 to find an analytical connection between them and corresponding period.
 
The Poncelet theorem, as one of the highlights of the XIX century projective
 geometry, attracted the attention of Arthur Cayley for several years
 (see [7-12]). In [8], Cayley found the analytical condition
 for caustic conics in the Euclidean plane case.
The classical and algebro-geometric proofs of  Cayley's theorem
can be found in Lebesgue's book [28] and Griffiths and Harris paper
[23], respectively.
The generalisation is established
by Dragovi\' c and Radnovi\' c for any $d$ [18, 19].
This
generalisation was done by use of the Veselov-Moser discrete quadratic
$L-A$ pair for the classical Heisenberg magnetic model [32].

The integrability of elliptical billiard systems in the Lobachevsky space was proved by
Veselov in [38]. There, Veselov used discrete linear $L-A$ pair,
which is quite different from the one used in the Euclidean case.

The starting point of this paper is derivation of Cayley's type
conditions for the Lobachevsky billiard and our observation that these
new conditions coincide with those obtained in [18, 19] for
the Euclidean case (Section 3).

We found a natural way to explain this coincidence and
it is related to the recently developed
integrability approach in the theory of geodesically equivalent
metrics [29, 35].
Both  Lobachevsky and Euclidean elliptic billiards can be
naturally considered as members of a hierarchy of integrable
elliptical billiards (Section 4). In the conclusion of this Section, we present
some properties of the Laurent polynomial integrable potential perturbations 
of those separable systems, continuing the study of such systems which
started with [14], see also [15, 24, 17, 16].
\medskip

\head{2. Basic notions on billiard systems}
\endhead

\medskip

Let $(Q,g)$ be a $d$--dimensional Riemannian manifold and let
$D\subset Q$ be a domain with a smooth boundary $\Gamma$.
Let $\pi: T^*Q \to Q$ be a natural projection and let $g^{-1}$
be the contravariant metric on the cotangent bundle,
in coordinates 
$$
\vert p\vert =\sqrt{g^{-1}(p,p)}=\sqrt{g^{ij}p_ip_j}, \quad
p\in T^*_x Q.
$$

Consider the {\it reflection mapping}
$$
r: \pi^{-1} \Gamma \to \pi^{-1} \Gamma,\quad p_- \mapsto p_+ ,
$$ 
which associates the covector
$p_+\in T^*_x Q$, $x\in \Gamma$ to a covector $p_-\in T^*_x Q$
such that  the  following conditions hold:
$$
\aligned
&\vert p_+ \vert =\vert p_- \vert, \\
&p_+-p_- \bot \Gamma.
\endaligned
\leqno(1)
$$

A {\it billiard} in $D$
is a dynamical system with the phase space
$M=T^*D$ whose trajectories are geodesics given by the Hamiltonian
equations
$$
\dot p=-\frac{\partial H}{\partial x}, \quad
\dot x=\frac{\partial H}{\partial p}, \quad H(p,x)=\frac12 g^{-1}_x(p,p),
\leqno(2)
$$
reflected at points $x\in \Gamma$ according to the billiard law:
$r(p_-)=p_+$. Here $p_-$ and $p_+$ denote the momenta before and after
the reflection.
If some potential force field $V(x)$ is added than
the system is described with the same reflection law (1)
and Hamiltonian equations (2) with the Hamiltonian
$H(p,x)=\frac12 g^{-1}_x(p,p)+V(x)$.

A function $f: T^*Q \to \Bbb{R}$ is an {\it integral} of the billiard
system if it commutes with the Hamiltonian ($\{f,H\}=0$) and
does not change under the reflection ($f(x,p)=f(x,r(p))$, $x\in\Gamma$).
The billiard is {\it completely integrable
in the sense of Birkhoff}
if it has $d$ integrals polynomial in the momenta, which are in involution,
and almost everywhere independent (see [26]).

The classical integrable examples, with smooth boundary,
are billiards inside ellipsoids
on the Euclidean and hyperbolic spaces and spheres, with
integrals quadratic in the velocities [26].
These systems can be also considered as 
discrete integrable systems [37, 38].
The explicit integrations
in terms of theta--functions
are performed by Veselov, Moser and Fedorov (see [37, 32, 38, 20]).

\medskip

\head{3. Poncelet theorem and Cayley's condition for the billiard
in the Lobachevsky space}
\endhead

\medskip

Veselov proved the integrability of the billiard
system within an ellipsoid in the Lobachevsky space in [38].
He showed that its motion
corresponds to certain translations of the Jacobi variety of some hyperelliptic
curve and gave explicit formulae of the motion in terms of theta-functions.
The aim of this section is to find  an analogue
of Poncelet's and Cayley's theorem [8] for the billiard motion within
an ellipsoid in the Lobachevsky space.

\medskip

\subhead{3.1. Integration of the billiard motion in the Lobachevsky space.
Poncelet theorem}
\endsubhead
For a brief account of Veselov's results
on the billiard in the Lobachevsky space [38],
let us consider the $(d+1)$-dimensional Minkowski space $V=\Bbb{R}^{d,1}$ with
the symmetric bilinear form:
$$
\langle \xi,\eta \rangle = -\xi_0\eta_0 + \xi_1\eta_1 + \dots +\xi_d\eta_d.
$$
One sheet of the hyperboloid $\langle \xi,\xi \rangle=-1$ with
the induced metric is a model of the $d$-dimensional Lobachevsky space
$\Bbb{H}^d$.
An ellipsoid $\Gamma$ in this space is determined by the equation
$$
\Gamma=\left\{\xi\in \Bbb{H}^d, \;
-{\xi_0^2\over a_0}+{\xi_1^2\over a_1}+\dots+{\xi_d^2\over a_d}
=0\right\},
\leqno(3)
$$
with $a_0>a_1 \ge a_2 \ge \dots \ge a_d>0$.

All segments of the billiard trajectory within this ellipsoid
are tangent to $d-1$ confocal quadric surfaces (including multiplicity),
fixed for a given trajectory (Theorem 3 in [38]).
Denote by $\mu_i,\; i=1,\dots , d-1$ the numbers such that the equations of these caustics are:
$$
-{x_0^2\over a_0-\mu_i}+{x_1^2\over a_1-\mu_i}+\dots+{x_d^2\over a_d-\mu_i}=0,
\quad (1\le i\le d-1).
\leqno(4)
$$

Then the points of reflection from the boundary $\Gamma$
correspond to the shift
$D_{k+1}= D_k + Q_- - Q_+$ on the Jacoby variety of the spectral
curve $\Cal C$
$$
\Cal C:\quad (\mu-a_0)\dots(\mu-a_d)=c\cdot \lambda^2 (\mu-\mu_1)\dots(\mu-\mu_{d-1}),
\leqno(5)
$$
where $c$ is a constant, and $Q_+,Q_-$ are the points on the curve
$\Cal C$ over $\mu=0$. (See Theorem 2 of [38].
The curve $\Cal C$ is the spectral curve
of the $L-A$ pair considered there.)

Let us note that Veselov considered only the case of
the regular (hyperelliptic) curve $\Cal C$ [38].
However, his consideration holds for the singular case, too.

Suppose a periodical billiard trajectory inside the
ellipsoid $\Gamma$ in the Lobachevsky space is given.
All trajectories with the same caustics have the same spectral curve.
If the period of the given trajectory is $n$,
then $n(Q_+ - Q_-)=0$ on $Jac(\Cal C)$, and vice-versa.      
Thus, all these trajectories close after $n$ bounces.
Therefore,     Poncelet's -type theorem for the billiard in the Lobachvsky space is derived from Veselov's results:

\proclaim{Proposition 1}
Suppose a periodical billiard trajectory inside an
ellipsoid in the Lobachevsky space is given.
Then any billiard trajectory which shares the same
caustic quadrics is also periodical, with the same period.
\endproclaim

\medskip

\subhead{3.2. Cayley's conditions --- regular spectral curve}
\endsubhead
Assume that all constants $a_0,a_1,\dots,a_d,\mu_1,\dots,\mu_{d-1}$ are
mutually different.
Then the spectral curve $\Cal C$ is hyperelliptic.
Cases when some of them coincide are discussed
in the next subsection.

To establish an analytical condition on a trajectory
to be periodic with period $n$, we need to find out when
the divisors $nQ_+$ and $nQ_-$ on the spectral curve are equivalent.

\proclaim{Lemma 1}
Let the curve $C$ be given by
$$
y^2=(x-x_1)\dots(x-x_{2g+2}),
\leqno(6)
$$
with all $x_i$ mutually different and not equal to $0$, and $Q_+$, $Q_-$ the two points on $C$ over the point $x=0$. Then $nQ_+\equiv nQ_-$ is equivalent to:
$$
\rank\left[\matrix
B_{g+2} & B_{g+3} & \dots & B_{n+1} \\
B_{g+3} & B_{g+4} & \dots & B_{n+2} \\
\dots & \dots & \dots & \dots \\
\\
\\
B_{g+n} & \dots & \dots & B_{2n-1}
\endmatrix\right]<n-g
\quad \text{and} \quad n>g,
\leqno(7)
$$
where $y=\sqrt{(x-x_1)\dots(x-x_{2g+2})}=B_0+B_1x+B_2x^2+\dots$ is the Taylor expansion around the point $Q_-$.
\endproclaim

\demo{Proof}
$C$ is a hyperelliptic curve of genus $g$. The relation $nQ_+\equiv nQ_-$ means that there exists a meromorphic function on $C$ with a pole of order $n$ at the point $Q_+$, a zero of the same order at $Q_-$ and neither other zeros nor poles. Denote by $L(nQ_+)$ the vector space of meromorphic functions on $C$ with a unique pole $Q_+$ of order at most $n$. Since $Q_+$ is not a branching point on the curve, $\dim L(nQ_+)=1$ for $n\le g$, and
$\dim L(nQ_+)=n-g+1$, for $n>g$. In the case $n\le g$,
the space $L(nQ_+)$ contains only constant functions,
and the divisors $nQ_+$ and $nQ_-$ can not be equivalent. If $n\ge g+1$, we choose the following basis for $L(nQ_+)$:
$$
1, f_1, \dots, f_{n-g},
$$
where
$$
f_k={y-B_0-B_1x-\dots-B_{g+k-1}x^{g+k-1}\over x^{g+k}}.
$$

Thus, $nQ_+\equiv nQ_-$ if there is a function $f\in L(nQ_+)$
with a zero of order $n$ at $Q_-$, i.e.,
if there exist constants $\alpha_0, \dots, \alpha_{n-g}$,
not all equal to 0, such that:
$$
\matrix
\alpha_0 & + & \alpha_1 f_1(Q_-) & + & \dots & \alpha_{n-g}f_{n-g}(Q_-) & = & 0 \\
 &  & \alpha_1 f_1'(Q_-) & + & \dots & \alpha_{n-g}f_{n-g}'(Q_-) & = & 0 \\
\dots \\
\\
\\
 &  & \alpha_1 f_1^{(n-1)}(Q_-) & + & \dots & \alpha_{n-g}f_{n-g}^{(n-1)}(Q_-) & = & 0.
\endmatrix
$$
Existence of a non-trivial solution to this system of linear equations is
equivalent to the condition (7).
\qed
\enddemo 

Introducing new coordinates $x=\mu$,
$y=\sqrt c \lambda (\mu-\mu_1)\dots(\mu-\mu_{d-1})$,
the spectral curve (5) is transformed to:
$$
y^2=(x-a_0)\dots(x-a_d)(x-\mu_1)\dots(x-\mu_{d-1}),
\leqno(8)
$$
and we obtain the following

\proclaim{Theorem 1}
The condition of a billiard trajectory inside the ellipsoid (3) in the
$d$-dimensional Lobachevsky space, with non-degenerate caustics (4),
 to be periodic with period $n\ge d$ is:
$$
\rank\left[\matrix
B_{n+1} & B_n     & \dots & B_{d+1} \\
B_{n+2} & B_{n+1} & \dots & B_{d+2} \\
\dots   & \dots   & \dots & \dots   \\
\\
B_{2n-1} & B_{2n-2} & \dots & B_{n+d-1}
\endmatrix\right]<n-d+1,
$$
where
$\sqrt{(x-a_0)\dots(x-a_d)(x-\mu_1)\dots(x-\mu_{d-1})}=B_0+B_1x+B_2x^2+\dots$.
There is no such trajectories with period less than $d$.
\endproclaim

\medskip

\subhead{3.3. Cases of singular spectral curve}
\endsubhead
When all $a_0, a_1, \dots, a_d, \mu_1, \dots, \mu_{d-1}$ are mutually
 different, then the curve (5) has no singularities in the affine part.
 However,  singularities appear in the following three cases
 and their combinations:

\smallskip

{\bf (i)}
$a_i=\mu_j$ for some $i,j$.
The spectral curve (5) decomposes into a rational and a hyperelliptic
curve. Geometrically,
this means that the caustic corresponding to
$\mu_i$ degenerates into hyper-plane $x_i=0$.
The billiard trajectory can be asymptotically tending
to that hyper-plane (and therefore can not be periodic),
or completely placed in this hyper-plane.
Therefore,
the closed trajectories appear when they are placed in a
coordinate hyper-plane.
Such motion can be discussed like in the case of dimension $d-1$.

\smallskip

{\bf (ii)}
$a_i=a_j$ for some $i\neq j$. The ellipsoid (3) is symmetric.

\smallskip

{\bf (iii)} $\mu_i=\mu_j$ for some $i\neq j$.
The billiard trajectory is placed on the corresponding confocal
quadric hyper-surface.
\footnote{We learned about its geometric significanse from Prof. Yurij Fedorov}

\smallskip

In the cases (ii) and (iii) the spectral curve $\Cal C$ is
a hyperelliptic curve with singularities.
In spite of their different
geometrical nature, they both need the same analysis of the condition
$nQ_+\equiv nQ_-$ for the singular curve (5).

\proclaim{Lemma 2}
Let the curve $C$ be given by
$
y^2=(x-x_1)\dots(x-x_{2g+2}),
$
with all $x_i$ different from $0$, and $Q_+$, $Q_-$ the two points on $C$
 over the point $x=0$. Then $nQ_+\equiv nQ_-$ is equivalent to $(7)$
where $y=\sqrt{(x-x_1)\dots(x-x_{2g+2})}=B_0+B_1x+B_2x^2+\dots$ is the Taylor
 expansion around the point $Q_-$.
\endproclaim

\demo{Proof}
Suppose that, among $x_1, \dots, x_{2g+2}$, only $x_{2g+1}$ and $x_{2g+2}$ have same values. Then $(x_{2g+1},0)$ is an ordinary double point on $C$. The normalisation of the curve $C$ is the pair $(\tilde C,\pi)$, where $\tilde C$ is the curve given by:
$$
\tilde C: \tilde y^2=(\tilde x-x_1)\dots(\tilde x-x_{2g}),
$$
and $\pi:\tilde C\to C$ is the projection:
$$
(\tilde x,\tilde y) \buildrel\pi\over\longmapsto
(x=\tilde x,\ y=(\tilde x-x_{2g+1})\tilde y).
$$
The genus of $\tilde C$ is $g-1$. The relation $nQ_+\equiv nQ_-$ is equivalent to existence of a meromorphic function $f$ on $\tilde C$,
$f\in L(n\tilde Q_+)$, with a zero of order $n$ at $\tilde Q_-$, and
$f(A)=f(B)$, where $\tilde Q_+$, $\tilde Q_-$ are the two points over
$\tilde x=0$, and $A,B$ are over $\tilde x=x_{2g+1}$.

For $n\le g-1$, $\dim L(n\tilde Q_+)=1$, and this space contains only constant functions. For $n\ge g$, we can choose the following basis for
$L(n\tilde Q_+)$:
$$
1, f_0, f_1 \circ \pi, \dots, f_{n-g}\circ\pi.
$$
$f_k$ are as in Lemma 1 for $k>0$, and
$$
f_0=
{\tilde y-\tilde B_0-\tilde B_1\tilde x-\dots-\tilde B_{g-1}\tilde x^{g-1}
\over
\tilde x^g}\,,
$$
where
$\tilde y=\sqrt{(\tilde x-x_1)\dots(\tilde x-x_{2g})}=
\tilde B_0+\tilde B_1\tilde x+\tilde B_2\tilde x^2+\dots$
is the Taylor expansion around the point $\tilde Q_-$.

Since $f_0$ is the only element of the basis with different values in the
 points $A$ and $B$, we obtain that $n\tilde Q_+\equiv n\tilde Q_-$ is
  equivalent to $(7)$.

Cases when $C$ has more singularities, or singular points of higher order,
can be discussed in the similar manner.
\qed
\enddemo

Immediate consequence of Lemma 2 is that Theorem 1 can be applied not
only for the case of the non-singular spectral curve, but in the cases
(ii) and (iii) too.
Threrefore, the following interesting property holds.

\proclaim{Theorem 2}
If the billiard trajectory within an ellipsoid $\Gamma$
in $d$-dimensional Loba\-chevsky space is
periodic with period $n<d$, then it is placed in one of the
$n$-dimensional planes of symmetry of the ellipsoid.
\endproclaim

This property can be seen easily for $d=3$.

\example{Example 1}
Consider the billiard motion in an ellipsoid in the
$3$-dimensional space, with $\mu_1=\mu_2$, when the segments
of the trajectory are placed on  generatrices of the corresponding
quadric surface confocal to the ellipsoid. If there existed a
periodic trajectory with period $n=d=3$, the three bounces would
have been complanar, and the intersection of that plane and the
quadric would have consisted of three lines, which is impossible.
It is obvious that any periodic trajectory with period $n=2$ is
placed along one of the axes of the ellipsoid. So, there is no
periodic trajectories contained in a confocal quadric surface,
with period less or equal to $3$.
\endexample

\medskip

\head{4. Hierarchy of integrable elliptical billiards}
\endhead

\medskip

\subhead{4.1. The Beltrami--Klein model of the Lobachevsky  space}
\endsubhead
Note first that Cayley's type
conditions for the Lobachevsky billiard from Section 3
are of the same form as those obtained in [18, 19] for
the Euclidean case, although the $L-A$ pairs used there
are quite different.
There is a natural way to explain this coincidence. 
We use the Beltrami--Klein model  of the
Lobachevsky  space $\Bbb{H}^d$.

The coordinate transformation
$$
y_1=\frac{\xi_1}{\xi_0},\dots,y_d=\frac{\xi_d}{\xi_0}
$$
maps  the Lobachevsky space, modeled as a pseudosphere of the Minkowski space,
to the Beltrami--Klein model within the unit sphere in $\Bbb{R}^d$ [38].

Now, after appropriate linear changing of
coordinates $x_1=\alpha_1 y_1,\dots,x_d=\alpha_d y_d$
we can obtain  the Beltrami--Klein model
inside the ellipsoid $\Lambda$:
$$
\Lambda=\left\{x=(x_1,\dots,x_d)\in \Bbb{R}^d, \;
\frac{x_1^2}{b_1}+\frac{x_2^2}{b_2}+\dots+\frac{x_d^2}{b_d} =1\right\},
$$
such that the ellipsoid $(3)$ in new coordinates is
confocal to $\Lambda$. Then its equation can be written
in the form:
$$
\Gamma=
\left\{x\in \Bbb{R}^d, \;
\frac{x_1^2}{b_1-c}+\frac{x_2^2}{b_2-c}+
\dots+\frac{x_d^2}{b_d-c} = 1\right\},
$$
where $0<c<b_i,\quad i=1,\dots, d$.

The hyperbolic metric within $\Lambda$ is given by
(for example, see [34]):
$$
\aligned
d\bar g^2=\frac{1}{b_1\cdot b_2 \dots b_d\cdot f^2} & \left( f
\left( \frac{dx_1^2}{b_1}+\dots+\frac{dx_d^2}{b_d}\right)+
\left( \frac{x_1 dx_1}{b_1}+\dots+\frac{x_d dx_d}{b_d}\right)^2\right),\\
&f=1-\left(\frac{x_1^2}{b_1}+\frac{x_2^2}{b_2}+\dots+\frac{x_d^2}{b_d}
\right).
\endaligned
$$

The metric $d\bar g^2$ can be written in the matrix form as
$d\bar g^2=\langle \Pi dx,dx\rangle $, where $dx=(dx_1,\dots,dx_d)$,
$$
\Pi=\frac{1}{\det\B\cdot f^2}
\left(f\B^{-1}+\B^{-1}x\otimes \B^{-1}x \right),
$$
$\B=\diag(b_1,\dots,b_d)$,
and $\langle \cdot,\cdot\rangle$ is the Euclidean scalar product.

The hyperbolic metric has
the same geodesics, considered as unparemetrized curves, as
 the Euclidean metric
$
dg^2=dx_1^2+dx_2^2+\dots+ dx_d^2.
$

Suppose that
$
b_1 > b_2 > \dots > b_d.
$
The standard elliptic coordinates $\lambda_1,\dots,\lambda_d$
in $\Bbb{R}^d$
($b_1>\lambda_1>b_2>\lambda_2>\dots>\lambda_{d-1}>b_d>\lambda_d$)
are defined
as solutions of the equation:
$$
\gamma(\lambda)=\frac{x_1^2}{b_1-\lambda}+\frac{x_2^2}{b_2-\lambda}+\dots+
\frac{x_d^2}{b_d-\lambda}=1.
$$

The direct verification shows that the metric $d\bar g^2$,
as well as the Euclidean metric $dg^2$, 
is orthogonally separable in the
elliptic coordinates and geodesic flows can be integrated by the
theorem of St\" ackel.
This means that hypersurfaces $\lambda_i=const$ of the
coordinate system $\lambda_1,\dots,\lambda_d$ are
orthogonal to each other and the corresponding Hamilton-Jacobi equations
for the Hamiltonian of the geodesic flows have
complete solutions of the form
$\Cal{S}(\lambda_1,\dots,\lambda_d,c_1,\dots,c_d)=
\Cal{S}_1(\lambda_1,c_1)+\dots+\Cal{S}_n(\lambda_d,c_d)$
(see [2, 3] and references therein).

Consider the billiard in the domain $D$
bounded by the ellipsoid $\Gamma$.
In elliptic coordinates, the boundary of the ellipsoid
is given by the equation $\lambda_d=c$ and the reflection
map, both for the Euclidean and Lobachevsky metrics, is given by
$$
\aligned
&(\lambda_1,\lambda_2,\dots,\lambda_{d-1},c,
p_{\lambda_1},p_{\lambda_2},\dots,p_{\lambda_{d-1}},p_{\lambda_d})
\mapsto\\
&\quad\quad(\lambda_1,\lambda_2,\dots,\lambda_{d-1},c,
p_{\lambda_1},p_{\lambda_2},\dots,p_{\lambda_{d-1}},-p_{\lambda_d}),
\endaligned
\leqno(9)
$$
where $(\lambda,p_\lambda)$ are canonical coordinates in $T^*\Bbb{R}^d$.
Such a simple form of the reflection map is due to the
fact that $\Gamma$ and $\Lambda$ are confocal in the coordinates
$x_1,\dots,x_d$.
Therefore we have

\proclaim{Lemma 3} The billiards inside the ellipsoid $\Gamma$ in
the Euclidean and the Lobachevsky space,
modeled within the ellipsoid $\Lambda$, have the
same trajectories up to reparametrization.
\endproclaim

Lemma 3 provides the explanation for the
coincidence of the Cayley's conditions obtained in the
previous section and papers [18, 19].
The above observation allows us to  approach
to the problem of the integrability
of elliptical billiards in a new way,
using theory
of geodesically equivalent metrics.

\medskip

\subhead{4.2. Geodesically equivalent metrics}
\endsubhead
Let $g$ and $\bar g$ be Riemannian metrics on $d$--dimensional
manifold $Q$.
The metrics $g$ and $\bar g$ are called {\it geodesically equivalent}
if they have the same geodesics considered as unparemetrized curves.
This is a classical subject 
studied by Beltrami, Dini, Levi-Civita, etc.
in 19th century. Recently, the new global unerstanding of the
theory is developed in the framework of integrable systems
(see [35, 36, 29, 5] and references therein).

Having the metrics $g$ and $\bar g$, define
the (1,1)--tensor field $\L=\L(g,\bar g)$ by
$$
\L=\left(\frac{\det(\bar g)}{\det(g)}\right)^{\frac{1}{d+1}}
\bar g^{-1} g.
$$

Consider functions
$$
J_l(p,x)=g^{-1}_x(\S_l p,p)=\sum_{j,k,i} {(\S_l)^i_j g^{jk}p_i p_k},
\leqno(10)
$$
where (1,1) tensors $\S_k$ are given by the formula:
$$
\{\det(\L+\alpha\Id)\}(\L+\alpha\Id)^{-1}=
\S_{d-1}\alpha^{d-1}+\S_{d-2} \alpha^{d-2} + \dots + \S_0,
\quad \alpha\in\Bbb{R}
$$

If the metrics $g$ and $\bar g$ are geodesically equivalent
then functions $J_l(p,x)$ are in involution with respect to the canonical
symplectic structure on $T^*Q$.
Moreover, if the eigenvalues of $\L$ are all different at one point of $Q$,
then they are different almost everywhere and the geodesic flows of $g$
and $\bar g$ are completely integrable. The complete set of involutive
integrals for the first flow is $J_0,J_1,\dots,J_{d-1}$ (see [35, 29, 36]).
In this case we say that $g$ and $\bar g$ are {\it strictly nonproportional}.

The pair $g$, $\bar g$ of geodesically equivalent Riemannian metrics produces
the family of geodesically equivalent Riemannian metrics $g_k$, $\bar g_k$,
$k\in \Bbb{Z}$ given by the formulas [35, 36]:
$$
g_k(\xi,\eta)=g(\L^k \xi,\eta), \quad \bar g_k(\xi,\eta)=g(\L^k\xi,\eta),
\quad k\in\Bbb{Z}.
\leqno(11)
$$
The integrals of the geodesic flows of metric $g_k$ and $\bar g_{k+2}$
canonically given by (10) coincide [35]. Following [5] we call (11)
{\it Topalov--Sinjukov hierarchy} of Riemannian metrics on $Q$.

The nice geometrical interpretation of geodesical equivalence
is done by Bolsinov and Matveev [5]. They proved
that $g$ and $\bar g$ are geodesically equivalent if and only
if $\L$ is a Benenti tensor field for
the metric $g$ [5, 3]. This implies that
if $g$ and $\bar g$ are strictly nonproportional than
all metrics $g_k$, $\bar g_k$ are orthogonally separable in the
same coordinates and geodesic flows can be integrated by the theorem
of St\" ackel [5].

\medskip

\subhead{4.3. Hierarchy of integrable elliptical billiards}
\endsubhead
Now we shall apply the general construction
described in the previous section to the
Euclidean $dg^2$ and Lobachevsky metrics $d\bar g^2$
inside the ellipsoid $\Lambda$.
Note that this natural geodesical equivalence is a slight modification
of the geodesical equivalence studied by Topalov in [36].
Taking $b=\B$, $\bar a=\sqrt{-1}x$, from lemma 7 of [36] we are getting
that (1,1) tensor field $\L$ has the following matrix form
$$
\L=\L(dg^2,d\bar g^2)=\left(\det\Pi\right)^{\frac{1}{d+1}}\Pi^{-1}=
\B-x\otimes x.
$$

Therefore we have the Topalov--Sinjukov hierarchy of
strictly  nonproportional Riemannian metrics within $\Lambda$ given by:
$$
\aligned
&dg_k^2=\langle (\B-x\otimes x)^k dx,dx \rangle, \\
& d\bar g_k^2= \langle \Pi (\B-x\otimes x)^k dx,
dx\rangle.
\endaligned
\leqno(12)
$$

The corresponding geodesic flows are completely integrable.
The integrals of the geodesic flow of the metrics
$dg_k^2$ and $d\bar g_{k+2}^2$
are:
$$
J^k_i(p,x)=\langle \S_i (\B-x \otimes x)^{-k} p, p \rangle
\leqno(13)
$$
where
$$
\aligned
&\{\det(\B-x\otimes x+\alpha\Id)\}(\B-x\otimes x+\alpha\Id)^{-1}=\\
&=\det\B_\alpha \left( (1-\langle \B_\alpha^{-1}x,x\rangle)
\B_\alpha^{-1}+\B_\alpha^{-1}\otimes \B_\alpha^{-1}\right)
=\S_{d-1}\alpha^{d-1}+\dots+\S_0,
\endaligned
\leqno(14)
$$
$p=(p_1,\dots,p_d)\in T^*_x\Bbb{R}^d$ is the
canonical momentum in Euclidean coordinates $x=(x_1,\dots,x_d)$,
$\B_\alpha=\diag(b_1+\alpha,\dots,b_d+\alpha)$ and
$\alpha$ is a real parameter.
For $k=0$ these functions are defined on the whole $T^*\Bbb{R}^d$ and
coincide with commuting functions given by Moser in [31].

Consider the billiards in the domain $D$ bounded by the ellipsoid
$\Gamma$ with metrics (12).
According to [5]
all   metrics (12) are orthogonally separable
in elliptical coordinates. This implies that the reflection map
is the same for all metrics and in elliptic coordinates
has the form (9). Moreover,
since integrals $J^k_i(p,x)$
are diagonal in elliptic coordinates, we have that
they are not just integrals
of the geodesic flows of metrics $dg_k^2$ and $d\bar g_{k+2}^2$,
but also integrals of the corresponding billiard systems inside ellipsoid
$\Gamma$ (see [26], pages 133-134).

Thus we get the following general statement:

\proclaim{Theorem 3}
The billiard systems inside ellipsoid $\Gamma$
with the Riemannian metrics $dg^2_k$, $d\bar g^2_k$ given by (12) are
completely integrable for all $k\in \Bbb{Z}$.
In particular, the elliptical
billiards in the Euclidean and hyperbolic spaces are completely integrable.
\endproclaim

Since the reflection map $r$ is the same for the whole
hierarchy,  applying Proposition 3 of [35], we get 

\proclaim{Corollary 1} The
billiard systems for the given hierarchy have isomorphic
Liouville foliations of $T^*D/r$.
\endproclaim

\remark{Remark 1}
Matveev and Topalov [29] and Tabachnikov [34] proved
that ellipsoid
$$
\tilde\Lambda=
\left\{x=(x_1,x_2,\dots,x_{d+1})\in \Bbb{R}^{d+1}, \;
\frac{x_1^2}{b_1}+\frac{x_2^2}{b_2}+\dots+\frac{x_{d+1}^2}{b_{d+1}} =1\right\}
$$
admits nontrivial geodesic equivalence between
the standard metric and
the metric
$$
\frac{1}
{b_1\cdot b_2 \dots b_{d+1}\cdot
\left(\frac{x_1^2}{b_1}+\frac{x_2^2}{b_2}+\dots+\frac{x_{d+1}^2}{b_{d+1}}
\right)}
\left( \frac{dx_1^2}{b_1}+\frac{dx_2^2}{b_2}+
\dots+\frac{dx_{d+1}^2}{b_{d+1}}\right)
\vert_{\tilde\Lambda}.
$$
The Euclidean and the Lobachevsky
metrics within ellipsoid $\Lambda$ can be seen as 
limits of the given metrics as $b_{d+1}$ tends to zero.
\endremark

\medskip

\subhead{4.4. Integrable potential perturbations}
\endsubhead
We shall say that the potential $V(x)$ is {\it separable in the
elliptic coordinates} $\lambda_1,\dots,\lambda_d$ if the
Hamilton--Jacobi equation for the Hamiltonian
$\frac12(p_1^2+\dots+p_d^2)+V(x)$ can be solved by separation of
variables in elliptic
coordinates.
This definition, in a more geometrical fashion, can be found in [2, 3].
The potential of the elastic force is an example.

The potential $V(x)$ is separable in the
elliptic coordinates on $\Bbb{R}^d$
if and only if $V(x)$ is a solution of the linear system
of partial differential equations
$$
(b_i-b_j)\frac{\partial^2 V}{\partial x_i \partial x_j}+
\left( x_i\frac{\partial}{\partial x_j}-  x_j\frac{\partial}{\partial x_i}\right)
\left(2V+\sum_{k=1}^d x_k  x_k\frac{\partial V}{\partial x_k}\right)=0,
\leqno(15)
$$
for $i\ne j$ (see [30, 2]).

We shall denote the Hamiltonians of the
geodesic flows of metrics $dg^2_k$ and $d\bar g^2_k$
by $H_0^k(p,x)$ and $\bar H_0^k(p,x)$ respectively.
Suppose that $V(x)$ is a solution of (15). Then from
[5] follows that Hamiltonian systems with
Hamiltonian functions
$$
H^k(p,x)=H_0^k(p,x)+V(x), \quad \bar H^k(p,x)=\bar H_0^k(p,x)+V(x)
\leqno(16)
$$
are completely integrable, and can be solved by separation of variables
in elliptic coordinates for all $k$.
There is a complete set of commuting integrals of the form
$$
I^k_i(p,x)=J^k_i(p,x)+f_i(x), \quad i=1,\dots,d,
$$
for each Hamiltonian $H^k(p,x)$ where $J^k_i(p,x)$ is given by (13).
The functions $f_i(x)$ do not depend of $k$. They are solutions of
the equations
$
\nabla f_i(x)=\S_i \nabla V(x),
$
where
$\nabla f=(\partial_1 f,\dots,\partial_n f)$ and $\S_i$ are given by (14).
Similar statement holds
for Hamiltonian systems with Hamiltonians $\bar H^k(p,x)$.

Consider the billiard systems with Hamiltonians (16)
within the ellipsoid $\Gamma$.
From the choice of $I^k_i(p,x)$ we have that these
functions do not change under the reflection.
Thus we get the following corollary.

\proclaim{Corollary 2}
Suppose that $V(x)$ is a solution of (15).
Then the billiard systems  with Hamiltonians (16)
within the ellipsoid $\Gamma$ are completely
integrable.
\endproclaim


Let us consider
the solution of equations (15) in the form of Laurent polynomials
$$
V(x)=\sum_{k_-\le i_1,\dots,i_d \le k_+} p_{i_1,\dots,i_d}
x_1^{i_1}x_2^{i_2}\dots x_d^{i_d}, \quad k_-,k_+ \in\Bbb{Z}.
\leqno(17)
$$

Suppose that Laurent polynomial (17) is a solution of (15).
Then coefficients $p_{i_1,\dots,i_d}$ satisfy
the following system of difference equations
$$
\aligned
&(b_{k}-b_{l})i_ki_l p_{i_1,\dots,i_d}= \\
&(i_1+\dots+i_d)
(i_l p_{i_1,\dots,i_{k-1},i_k-2,i_{k+1},\dots,i_d}-
i_k  p_{i_1,\dots,i_{l-1},i_l-2,i_{l+1},\dots,i_d}).
\endaligned
\leqno(18)
$$

Such potential perturbations are described for $d=2$ in [14] (see also
[15, 24]) and for $d=3$ in [17].
In general, 
the linear space of 
Laurent polynomial solution of (15) has a basis of the form
$$
\aligned
&\Cal{V}_k=\Cal{V}_k(x_1,\dots,x_d), \\
&\Cal{W}_k^{i}=\frac{1}{x_i^{2k}}\Cal{P}_{k-1}^{i}(x_1,\dots,x_d),
\quad i=1,\dots,d,
\endaligned
\leqno(19)
$$
where $\Cal{V}_k$ and $\Cal{P}_k^i$ are polynomials of
degree $2k$, $k\ge 0$.
The potentials $\Cal{V}_k$ and $\Cal{W}_k^i$, in elliptic coordinates,
correspond to the potentials 
$$
V(\lambda)=\sum_{j=1}^d\frac{v(\lambda_j)}{\Pi_{l\ne j}(\lambda_j-\lambda_l)},
$$
with $v(t)=\sum_{j=1}^k \alpha_j {t^{d-1+j}}$ and
$v(t)= \sum_{j=1}^k \beta_j (t-a_i)^{-j}$ respectively.

\example{Example 2}
As an example,
we write down a few of the basis potentials (19):
$$
\aligned
&\Cal{V}_1(x)=\sum_j x_j^2 \; (\text{Jacobi}),
\quad
\Cal{V}_2(x)=\sum_j b_jx_j^2-\left(\sum_j x_j^2\right)^2,\\
&\Cal{V}_3(x)=\sum_j b_j^2x_j^2-
2\left(\sum_j x_j^2\right)^2\left(\sum_j b_jx_j^2\right)
+ \left(\sum_j x_j^2\right)^3, \\
&\Cal{W}_1^i(x)=\frac{1}{x_i^2}\; (\text{Rosochatius}),
\quad
\Cal{W}_2^i(x)=
\frac{1}{x_i^4}
\left(1+\sum_{j\ne i}\frac{x_j^2}{b_i-b_j}\right),\\
&\Cal{W}^i_3(x)=\frac{1}{x_i^6}\left(
1+\sum_{j\ne i}\left(
\frac{2x_j^2}{b_i-b_j}+
\frac{x_j^2x_i^2}{(b_i-b_j)^2}\right)+
\sum_{j,k \ne i} \frac{x_j^2 x_k^2}{(b_i-b_j)^2(b_i-b_k)^2}\right).
\endaligned
$$
\endexample

Let $\Cal{V}(x)=\sum_p\alpha_p \Cal{V}_p(x)$
be some separable polynomial potential.
Consider billiard systems with Hamiltonians (16).
By the Maupertiues principle [1],
for a given value of total energies $h$,
satisfying condition $h>\max_{x\in D} \Cal{V}(x)$,
the motions in the potential field $\Cal{V}(x)$ inside $\Gamma$
are reduced to geodesical motions with metrics
$$
(h-\Cal{V}(x))dg^2_k, \quad(h-\Cal{V}(x))d\bar g^2_k.
\leqno(20)
$$
It is clear that the billiard systems within $\Gamma$
with metrics (20) are integrable.

\remark{Concluding remarks}
Theorem 3 holds also if the boundary of the billiard is the
union of the confocal quadrics
$\Gamma=\Gamma_{c_1}\cup\Gamma_{c_2}\dots\cup \Gamma_{c_r}$,
$\Gamma_{c_i}=
\{x\in\Bbb{R}^d, \; \gamma(c_i)=1\}$,
or more generally, if the billiard is constrained to some of confocal
quadrics.
The same results can be formulated for billiards constrained on
spheres by using  geodesical equivalences
established in [35, 29]. Then systems are orthogonally separable
in  the spherical elliptic coordinates.

Polynomial potentials separable in elliptic
coordinates on $\Bbb{R}^d$ and spheres $S^d$
are given by Bogoyavlenski [4] and  Wojceichowski [39].
After Rosochatius's potential $V(x)=\sum_{i=1}^d{\alpha_i}x_i^{-2}$
(see [33] and Apendix in [31]),
particular examples of rational potentials
are found by Braden and Wojceichowski [6, 39].
Kalnins, Benenti and Miller described separable rational potentials in terms
of certain recurrence relations between potentials
of different degrees [25]. 

Recently, Fedorov integrated the elliptical billiard with
the elastic potential in the Euclidean spase [21].
Dragovi\' c found the connection between the
Laurent potential perturbations of the elliptical billiards for $d=2$ and 
the Appell hypergeometric functions [16].
Also recently, the two--dimensional billiards with smooth boundary
and additional irreducible integrals of the third and fourth degree,
with a help of the integrable cases of Goryachev--Chaplygin and
Kovalevskaya are given by Kozlova [27].
\endremark

\medskip

\subhead{Acknowledgement}
\endsubhead
The research was partially supported by
the Serbian Ministry of Science
and Technology, Project 1643 --
Geometry and Topology of Manifolds and
Integrable Dynamical Systems.
The research of one of the authors (V.D.) was also partially supported by SISSA
and MURST Project Geometry of Integrable Systems.

\medskip

\head{References}
\endhead

\medskip

\item{[1]}
Arnol'd, V.:
{\it Mathematical methods of classical mechanics},
Springer, 1978.

\item{[2]}
Benenti, S.:
Orthogonal separable dynamical systems.
In: {\it Differential geometry and its applications (Opava, 1992)},
163-184, Math. Publ. 1,
Silesian Univ. Opava (1993);
Electronic edition: http://www.emis.de/proceedings/.

\item{[3]}
Benenti, S.:
Inertia tensors and St\" akel systems in Euclidean space,
Rend. Sem. Mat. Univ. di Torino {\bf 50} (1992) no. 4, 315-341.

\item{[4]}
Bogoyavlensky, O. I.:
Integrable cases of a rigid body dynamics and integrable systems
 on spheres $S^n$,
Izv. Akad. Nauk SSSR Ser. Mat. {\bf 49} (1985) no. 5, 899-915 (in Russian);
English translation: Math. USSR-Izv {\bf 27} (1986) no. 2, 203-218.

\item{[5]}
Bolsinov, A. V. and Matveev, V. S.:
Geometrical interpretation of Benenti systems,
to appear in Journal of Geometry and Physics.

\item{[6]}
Braden, H. W.:
A completely integrable mechanical system,
Lett. Math. Phys. {\bf 6} (1982) 449-452.

\item {[7]} Cayley, A.: Note on the porism of the in-and-circumscribed polygon,
Philosophical magazine, vol. VI (1853), pp. 99-102
 
\item {[8]} Cayley, A.: Developments on the porism of the in-and-circumscribed polygon,
Philosophical magazine, vol. VII (1854), pp. 339-345
 
\item {[9]} Cayley, A.: On the porism of the in-and-circumscribed  triangle, and on an irrational
transformation of two ternary quadratic forms each into itself,
Philosophical magazine, vol. IX (1855), pp. 513-517
 
\item {[10]} Cayley, A.: On the porism of the in-and-circumscribed  triangle,
Quarterly Mathematical Journal, vol. I (1857), pp. 344-354
 
\item {[11]} Cayley, A.: On the a posteriori demonstration of the porism of the in-and-circum\-scribed
triangle,
Quarterly Mathematical Journal, vol. II (1858), pp. 31-38

\item {[12]} Cayley, A.: On the porism of the in-and-circumscribed polygon,
Philosophical Transactions of the Royal Society of London, vol. CLI (1861),
pp. 225-239

\item{[13]}
Chang, S.-J., Crespi, B. and Shi, K.-J.:
Elliptical billiard systems and the full Poncelet's theorem in $n$ dimensions,
J. Math. Phys. {\bf 34} (1993), 2242-2256

\item{[14]}
Dragovi\' c, V.:
On integrable potentials of the billiard system within ellipse,
Prikl. Math. Mech. {\bf 62} (1998) no. 1, 166-169 (Russian);
English translation: J. Appl. Math. Mech. {\bf 62} (1998) no. 1, 159-192.

\item{[15]}
Dragovi\' c, V.:
On integrable potential perturbations of the Jacobi problem
for the geodesic on the ellipsoid,
J. Phys. A: Math. Gen. {\bf 29} (1996) L317-L321.

\item{[16]}
Dragovi\' c, V.: The Appell hypergeometric functions and classical separable 
mechanical systems, J. Phys A: Math. Gen. {\bf 35} (2002) 2213-2221

\item{[17]}
Dragovi\' c, V. and Jovanovi\' c, B.:
On integrable potential perturbations of the billiard systems
within an ellipsoid,
J. Math. Phys. {\bf 38} (1997) 3063-3068.

\item{[18]}
Dragovi\' c, V. and Radnovi\' c, M.:
Conditions of Cayley's type for ellipsoidal billiard,
J. Math. Phys. {\bf 39} (1998) 355-362.

\item{[19]}
Dragovi\' c, V. and Radnovi\' c, M.:
On periodical trajectories of the billiard systems within an
ellipsoid in $\Bbb{R}^d$ and generalized Cayley's condition,
J. Math. Phys. {\bf 39} (1998) 5866-5869.

\item{[20]}
Fedorov, Yu.:
Classical integrable systems and billiards related to generalized Jacobians,
Acta Appl. Math. {\bf 55} (1999) 251-301.

\item{[21]}
Fedorov, Yu.:
An ellipsoidal billiard with a quadratic potential,
Funct. Anal. Appl. {\bf 35} (2001) no. 3, 199-208.

\item{[22]} Griffiths, P. and Harris, J.:
A Poncelet theorem in space,
Comm. Math. Helv. {\bf 52} (1977) 145-160.

\item{[23]} Griffiths, P. and Harris, J.:
On Cayley's explicit solution to Poncelet's porism,
Enseignement Math. {\bf 24} (1978) 31-40.

\item{[24]}
Jovanovi\' c, B.:
Integrable perturbations of billiards on costant curvature surfaces,
Phys. Lett. A {\bf 221} (1997) 353-358.

\item{[25]}
Kalnins, E. G., Benenti, S. and Miller, W.:
Integrability, St\" akel spaces, and rational potentials,
J. Math. Phys. {\bf 38} (1997) 2345-2365.

\item{[26]}
Kozlov, V. V. and Tresch\" ev, D. V.:
{\it Billiards}, Transl. of Math. Monographs, Amer. Math. Soc. Volume 89,
Providence RI, 1991.

\item{[27]}
Kozlova, T.:
Billiard systems with polynomial integrals of third and fourth degree,
J. Phys. A: Math. Gen. {\bf 34} (2001) 2121-2124.

\item{[28]}
Lebesgue, H.:
{\it Les coniques},
Guthier-Villars, Paris, 1972.

\item{[29]}
Matveev, V. S. and Topalov, P.:
Quantum integrability of Beltrami--Laplace operator as geodesic equaivalence.
Math. Z. {\bf 238} (2001) 833-866. 

\item{[30]}
Marshall, I., Wojciechowski, S.:
When is a Hamiltonian system separable?
J. of Math. Phys., {\bf 29} (1988) 1338-1346.

\item{[31]}
Moser, J.:
Geometry of quadrics and spectral theory.
In: {\it The Chern Symposium 1979}, pp. 147-188, Springer, New York-Berlin, 1980.

\item{[32]}
Moser, J. and Veselov, A. P.:
Discrete version of some classical integrable systems and factorization of
matrix polynomials,
Comm. Math. Phys. {\bf 139} (1991) 217-243.

\item{[33]}
Rosochatius, E.:
 Uber die Bewegung eines Puktes, Inaugural Dissertation. Univ. Gottingen. Gebr. Unger. Berlin (1877)

\item{[34]}
Tabachnikov, S.:
Projectively equivalent metrics, exact transverse line field,
and the geodesic flow on the ellipsoid,
Comm. Math. Helv. {\bf 74} (1999) 306-321.

\item{[35]}
Topalov, P.:
Integrability criterion of geodesical equaivalence. Hierarchies,
Acta Appl. Math. {\bf 59} (1999) 271-298.

\item{[36]}
Topalov, P.:
Geodesic hierarchies and involutivity,
J. Math. Phys. {\bf 42} (2001) 3898-3914.

\item{[37]}
Veselov, A. P.:
Integrable systems with discrete time and difference operators,
Fuct. Anal. Appl. {\bf 22} (1988) no. 2, 1-13 (Russian);
English translation: Funct. Anal. Appl. {\bf 22} (1998) no. 2, 83-93.

\item{[38]}
Veselov, A. P.:
Confocal surfaces and integrable billiards on the sphere and in the
Lobachevsky space,
J. Geom. Phys. {\bf 7} (1990) no. 1, 81-107.

\item{[39]}
Wojciechowski, S,:
Integrable one-partical potentials related to the Neumann system and
the Jacobi problem of geodesic motion on an ellipsoid,
Phys. Lett. A {\bf 107} (1985) 107-111.

\end

\remark{Remark}
Matveev and Topalov [MaTo] and Tabachnikov [Ta] proved
that ellipsoid $E^{d-1}$ admits
geodesically equivalent metrics, standard one
$
ds^2=dx_1^2+\dots+dx_{d}^2\vert_{E^{d-1}}
$
and
$$
d\bar s^2=\frac{1}
{b_1\cdot b_2 \dots b_d\cdot
\left(\frac{x_1^2}{b_1}+\frac{x_2^2}{b_2}+\dots+\frac{x_d^2}{b_d}\right)}
\left( \frac{dx_1^2}{b_1}+\dots+\frac{dx_{d}^2}{b_d}\right)\vert_{E^{d-1}}.
$$
There is an interesting duality between the above metrics and
the Euclidean and
Beltrami--Klein metrics.
Metrics $dg^2$ and $d\bar g^2$ on $Q^{d-1}\subset \Bbb{R}^{d-1}$
can be seen as a limits
of metrics $ds^2$ and $d\bar s^2$ as $a_{d}$ tends to zero.
On the other side, Tabachnikov showed that the metric $d\bar s^2$
can be gotten, after appropriate "renormalization",
from the Beltrami--Klein metric within the ellipsoid $E^{d-1}$
(see theorem 3.2 in [Ta]).
\endremark

\remark{Remark}
The theorem holds also if the boundary of the billiard is the
union of the confocal quadric
$\Gamma=\Gamma_{c_1}\cup\Gamma_{c_2}\dots\cup \Gamma_{c_r}$,
$\Gamma_{c_i}=
\{x\in\Bbb{R}^n, \; \gamma(c_i)=1\}$,
or more generally, if the billiard is constrained to some of confocal
quadric.
The same result can be formulated for billiards constrained on
spheres by using  geodesical equivalences
established in [MaTo]. Then systems are orthogonally separable
in  the spherical elliptic coordinates
$\lambda_1,\dots,\lambda_{d-1}$ on $S^{d-1}$.
\endremark

The complete set of commuting integrals can be taken
to be of the form (for Hamiltonians $H^k(p,x)$):
$$
I^k_i(p,x)=J^k_i(p,x)+f_i(x), \quad i=1,\dots,n,
$$
where $J^k_i(p,x)$ are given by (13).
The functions $f_i(x)$ do not depend of $k$ and they are solution of
the equations
$$
\nabla f_i(x)=\S_i \nabla V(x),
$$
where
$\nabla f=(\partial_1 f,\dots,\partial_n f)$ and $\S_i$ are given by (14).
Similar statement holds
for Hamiltonian systems with Hamiltonians $\bar H^k(p,x)$.

\remark{Remark}
Polynomial potentials that are separable in elliptic
coordinates on $\Bbb{R}^d$ and spheres $S^d$
are given by Bogoyavlenski [Bo] and  Wojceichowski [Wo].
After Rosochatius's potential $V(x)=\sum_{i=1}^d{\alpha_i}x_i^{-2}$
(see apendix in [Mo]),
particular examples of rational potentials
are found by Braden
($V(x)=\left(\sum_{i=1}^d x_i^2/a_i\right)^{-1}$, separable on $S^d$)
and Wojceichowski
($V(x)=\left(1-\sum_{i=1}^d x_i^2/a_i\right)^{-1}$, separable on
$\Bbb{R}^d$) [Br, Wo].
Kalnins, Benenti and Miller described separable rational potentials in terms
of certain recurrence relations between potentials
of different degrees [KaBeMi].
\endremark

Suppose that $\Gamma$
consists of a
finite number of smooth, regular hypersurfaces $\Gamma_i$
that intersect transversally.
If all two-faced angles are integral parts of $\pi$ (i.e, are equal to
$\pi/m$, $m=2,3,\dots$) then we also have well
defined reflection map (see [KoTr]).

&\Cal{W}_3^i(x)=\frac{1}{x_i^6}
\left(1+\frac{2x_1^2}{b_i-b_1}+
\dots+\frac{2x_{i-1}^2}{b_i-b_{i-1}}+
\frac{2x_{i+1}^2}{a_i-a_{i+1}}+\dots+
\frac{2x_d^2}{b_i-b_d}
\right)+\\
&\quad\frac{1}{x_i^6}
\left(\frac{2x_1^2x_2^2}{(b_i-b_1)(b_i-b_2)}+
\frac{2x_{1}^2x_3^2}{(b_i-b_{1})(b_i-b_3)}+\dots+
\frac{2x_{d-1}^2 x_d^2}{(b_i-b_{d-1})(b_i-b_d)}
\right)+\\
&\quad\frac{1}{x_i^6}
\left(\frac{x_1^2x_i^2}{(b_i-b_1)^2}+
\dots+\frac{x_{i-1}^2x_i^2}{(b_i-b_{i-1})^2}+
\frac{x_{i+1}^2x_i^2}{(b_i-b_{i+1})^2}+\dots+
\frac{x_d^2x_i^2}{(b_i-b_d)^2}
\right)+\\
&\quad\frac{1}{x_i^6}
\left(\frac{x_1^4}{(b_i-b_1)^2}+
\dots+\frac{x_{i-1}^4}{(b_i-b_{i-1})^2}+
\frac{x_{i+1}^4}{(b_i-b_{i+1})^2}+\dots+
\frac{x_d^4}{(b_i-b_d)^2}
\right).\\